\documentclass[
    reprint,
    superscriptaddress,
    amsmath,amssymb,
    aps,
    prx,
    longbibliography,
]{revtex4-2}
\usepackage{braket}
\usepackage{graphicx}
\usepackage{amsmath}
\usepackage{amssymb}
\usepackage{algorithm}
\usepackage{algpseudocode}
\usepackage{bbm}
\usepackage{multirow}
\usepackage{xcolor}
\usepackage[english]{babel}
\usepackage[colorlinks=true,urlcolor=blue,linkcolor=blue,citecolor=blue]{hyperref}
\usepackage[margin=0.75in]{geometry}

\newcommand{\Tr}{\operatorname{Tr}}
\newcommand{\eff}{ \mbox{\tiny eff}}
\newcommand{\SF}{ \mbox{\tiny SF}}
\newcommand{\TFI}{ \mbox{\tiny TFI}}
\newcommand{\GS}{ \mbox{\tiny GS}}
\newcommand{\MPS}{ \mbox{\tiny MPS}}
\newcommand{\unity}{\mathbbm{1}}
\newcommand{\Eq}[1]{Eq.\@ \ref{#1}}
\newcommand{\Fig}[1]{Fig.\@ \ref{#1}}

\definecolor{LBlue}{rgb}{0,0.34,0.45}

\begin{document}

\title{Large-scale Density Matrix Renormalization Group simulations on accelerator clusters}

\title{Density Matrix Renormalization Group with Tensor Processing Units}

\author{Martin Ganahl}
\affiliation{SandboxAQ, Palo Alto, CA, USA}
\affiliation{Sandbox@Alphabet, Mountain View, CA 94043, USA}

\author{Jackson Beall}
\affiliation{SandboxAQ, Palo Alto, CA, USA}
\affiliation{Sandbox@Alphabet, Mountain View, CA 94043, USA}

\author{Markus Hauru}
\affiliation{Sandbox@Alphabet, Mountain View, CA 94043, USA}
\affiliation{The Alan Turing Institute, 96 Euston Road, London, England, NW1 2DB, UK}

\author{Adam G.M. Lewis}
\affiliation{SandboxAQ, Palo Alto, CA, USA}
\affiliation{Sandbox@Alphabet, Mountain View, CA 94043, USA}

\author{Jae Hyeon Yoo}
\affiliation{Sandbox@Alphabet, Mountain View, CA 94043, USA}
\affiliation{X, the Moonshot Factory, Mountain View, CA 94043, USA}
\affiliation{Google Core, Mountain View, CA 94043, USA}

\author {Yijian Zou}
\affiliation{Sandbox@Alphabet, Mountain View, CA 94043, USA}
\affiliation{Stanford Institute for Theoretical Physics, Stanford University, Palo Alto, CA 94305, USA}

\author{Guifre Vidal}
\affiliation{Sandbox@Alphabet, Mountain View, CA 94043, USA}
\affiliation{X, the Moonshot Factory, Mountain View, CA 94043, USA}
\affiliation{Google Quantum AI, Mountain View, CA 94043, USA}

\date{\today}

\begin{abstract}
Google's Tensor Processing Units (TPUs) are integrated circuits specifically built to accelerate and scale up machine learning workloads. They can perform fast distributed matrix multiplications and therefore be repurposed for other computationally intensive tasks. In this work we demonstrate the use of TPUs for accelerating and scaling up the density matrix renormalization group (DMRG), a powerful numerical approach to compute the ground state of a local quantum many-body Hamiltonian. The cost of DMRG scales with system size $N$ as $O(ND^3)$, where the so-called bond dimension $D$ regulates how expressive the underlying matrix product state (MPS) variational ansatz is. We consider lattice models in two spatial dimensions, with square lattices of size $10\times 10$ (free fermions) and $20\times 20$ (transverse field Ising model), for which the required MPS bond dimension is known to scale at least as $\exp(\sqrt{N})$. Using half of a TPU v3 pod (namely $1,\!024$ TPU v3 cores) we reached an unprecedentedly large bond dimension $D = 2^{16} = 65,\!536$, for which optimizing a single MPS tensor took about 2 minutes.
\end{abstract}

\maketitle
\section{Introduction}
The Density Matrix Renormalization Group (DMRG) \cite{white_density_1992, white_density-matrix_1993} algorithm is the gold standard method for computing ground-states and low-lying excited states of one-dimensional (1d) local Hamiltonians \cite{hallberg_density-matrix_1995, kuhner_dynamical_1999,jeckelmann_dynamical_2002, barthel_spectral_2009, pirvu_matrix_2012, haegeman_variational_2012}. Since its original formulation, the DMRG method and its descendants \cite{nishino_density_1995, nishino_corner_1996, vidal_efficient_2003, verstraete_renormalization_2004, vidal_entanglement_2007, vidal_class_2008, mcculloch_infinite_2008} have been applied to a wide variety of problems, both in one and higher dimensions, ranging from quantum chemistry \cite{white_ab_1999, chan_density_2011, wouters_density_2014,szalay_tensor_2015, reiher_dmrg_nodate, olivares-amaya_ab-initio_2015, yanai_density_2015,chan_matrix_2016, white_multi-sliced_2019, baiardi_density_2020,brabec_massively_2020,barcza_dmrg_2020, qiu_hybrid_2021,goings_reliably_2022}
to material science \cite{ganahl_efficient_2015,ganahl_chebyshev_2014, bauernfeind_fork_2017}, quantum computing \cite{hastings_area_2007, verstraete_matrix_2008, huang_efficient_2021, bravyi_efficient_2014, ferris_tensor_2014, chubb_statistical_2021, darmawan_linear-time_2018, markov_simulating_2008, boixo_simulation_2018} and machine learning \cite{gao_quantum_nodate, cheng_tree_2019, stoudenmire_supervised_2017, stoudenmire_learning_2018, glasser_probabilistic_2019, vieijra_generative_2022, roberts_tensornetwork_2019, han_unsupervised_2018}. Tensor network methods, of which DMRG is the most successful incarnation, hold the promise of revolutionizing these fields. 

DMRG is an optimization method over a matrix product state (MPS) \cite{fannes_finitely_1992, affleck_rigorous_1987, ostlund_thermodynamic_1995, rommer_class_1997, dukelsky_equivalence_1998}, which is a powerful variational ansatz in the form of a one-dimensional tensor network. For a system size $N$ (where $N$ denotes e.g. the number of sites in a lattice model or the number of electronic orbitals in a molecule) the MPS is made of $N$ tensors and the computational cost of DMRG scales as $O(ND^3)$. Here, the so-called bond dimension $D$ determines how much quantum entanglement the MPS is capable of accounting for, and may depend on the system size, that is $D=D(N)$. For instance, to accurately represent a generic wavefunction, the bond dimension must grow as $D = \exp(N)$, making the MPS optimization as expensive as a direct, brute-force ground state computation. Fortunately, most ground states of local Hamiltonians in $d$ spatial dimension contain restricted amounts of entanglement according to the so-called \textit{area law} (with possible logarithmic corrections). Both the incredible success of DMRG for one-dimensional systems and its challenges in higher dimensions can be understood to be a direct consequence of this area law.

Such applications of DMRG in $d>1$ dimensions, while  exceedingly difficult and computationally expensive, are also tremendously important in order to account for exotic quantum effects that other, less expensive methods fail to capture. Recent years have seen a growing effort to understand how modern computer architectures, in particular HPC clusters, can be used to speed up such computations \cite{zhai_low_2021, motoyama_tenes_2021, brabec_massively_2020, levy_distributed-memory_2020, ueda_corner_2020, nemes_density_2014}. 
In this work we show that Googles's Tensor Processing Units (TPUs) \cite{TPUinfo, jouppi2017datacenter}, originally developed for machine learning workloads but more recently applied to other computational tasks \cite{tpu_floquet, tpu_qphys, tpu_algebra, tpu_qchem, tpu_qhardware, tpu_Z2field, tpu_circuit, tpu_matsci, tpu_proteins}, can be leveraged to perform, within hours, large scale DMRG calculations of 2d quantum systems that would otherwise take many months to finish on conventional shared memory hardware with up to a few dozens of CPU cores. We demonstrate the approach in 2d square lattice models of sizes $10\times 10$ and $20 \times 20$. To the best of our knowledge, the largest bond dimension employed, $D = 2^{16} = 65,\!536$, sets a new record. (This was achieved using only 1024 TPU v3 cores, namely half of a TPU v3 pod, and without exploiting internal symmetries in the MPS representation, and therefore there is significant room for further increasing $D$, see discussion section). These results herald a new age of DMRG and, more generally, tensor network methods, with the potential to transform the computational landscape in all research areas where such techniques are applied, from condensed matter to quantum chemistry, materials science and machine learning. 

The paper is organized as follows: in Sect. \ref{sec: dmrg} we review some relevant aspects of the DMRG algorithm, the MPS ansatz and the entanglement area law; in Sect. \ref{sec: tpus} we then briefly describe TPUs; in Sect. \ref{sec:dmrg on tpus} we introduce the strategy used to distribute DMRG on TPUs, including data distribution, a necessary out-of-core approach, and distributed tensor contractions; in Sect. \ref{sec: results} we present benchmark results using two models on a square lattice: free fermions and the transverse field Ising model; we conclude the paper with a summary and discussion in Sect. \ref{sec: discussion}. We also include Appendices \ref{sec: ortho} - \ref{sec:subspace} with additional technical details of our DMRG implementation.

\section{Density Matrix Renormalization Group}
\label{sec: dmrg}

In this section we present a brief review of the Density Matrix Renormalization Group (DMRG) algorithm \cite{white_density_1992} and the matrix product state (MPS) ansatz \cite{fannes_finitely_1992}, as well as other relevant background material. 

We consider a lattice system made of $N$ sites, with each site described by a vector space of finite dimension $q$ and orthonormal basis $\{\ket{i}\}$, $i=1,2,\cdots,q$. For instance, with $q=2$, each site is represented by a two-dimensional vector space with orthonormal basis $\{\ket{1},\ket{2}\}$, corresponding e.g. to empty/occupied fermionic states $\{\ket{0}, \ket{1}\}$ if each site represents a spinless fermionic degree of freedom, or to spin up/spin down states $\{\ket{\uparrow}, \ket{\downarrow}\}$ if each site represents a spin-$\frac{1}{2}$ quantum spin degree of freedom, as in the two examples used later on in this paper. The many-body wavefunction of the lattice system then reads
\begin{equation} \label{eq:psi}
    \ket{\psi} = \sum_{i_1i_2\cdots i_N} \psi^{i_1i_2\cdots i_N} \ket{i_1i_2\cdots i_N},
\end{equation}
where $\psi^{i_1i_2\cdots i_N}$ denotes $q^N$ (possibly complex) amplitudes and $\ket{i_1i_2\cdots i_N}$ stands for the product basis $\ket{i_1} \otimes \ket{i_2} \otimes \cdots \otimes \ket{i_N}$ for the $q^N$-dimensional vector space of the $N$ sites. Similarly, the local many-body Hamiltonian expressed in the same basis reads
\begin{equation} \label{eq:H}
   \mathcal{H} = \sum_{\{i\},\{j\}} \mathcal{H}^{i_1j_1i_2j_2\cdots i_Nj_N} \ket{i_1i_2\cdots i_N}\bra{j_1j_2\cdots j_N},
\end{equation}
although a more natural, efficient expression is as a sum of local terms. Our goal is to compute an accurate approximation to the ground state $\ket{\psi_{\GS}}$ of the lattice Hamiltonian $\mathcal{H}$, without explicitly storing the wavefunction amplitudes in \eqref{eq:psi}, which would incur in a computational cost exponential in the system size $N$.

\subsection{Matrix Product Decompositions}

For that purpose, one can use the Density matrix Renormalization Group (DMRG) algorithm \cite{white_density_1992}, which is a variational method in the space of matrix product states (MPSs) \cite{fannes_finitely_1992}. The MPS ansatz consists of a collection of N order-3 tensors
\begin{equation} \label{eq:MPS}
\{M_1, M_2, ~\cdots, ~M_N\}.
\end{equation}
Each tensor $M_k$ has (possibly complex) components $[M_k]_{\alpha_{k-1} \alpha_{k} }^{i_k}$, where each index $\alpha_k$ takes $D_k$ different values (that is, $\alpha_{k} = 1,2, \cdots, D_k$). In other words, for each value of $i_k \in \{1,2,\cdots, q\}$, tensor $M_k$ defines a matrix $[M_k]^{i_k}$ of size $D_{k-1}\times D_{k}$ with matrix elements labeled by indices $\alpha_{k-1}$ and $\alpha_k$. Following a common practice, we will refer to the index $i_k$ labeling the local basis of states as a \textit{physical index}, and to the indices $\alpha_{k-1}$ and $\alpha_k$ as \textit{bond indices}. The bond indices $\alpha_0$ and $\alpha_N$ will be chosen to have dimensions $D_0 = D_N = 1$, so that for fixed value of the physical indices $i_1$ and $i_N$, $[M_1]^{i_1}$ and $[M_N]^{i_N}$ are not matrices but vectors (of dimension $D_1$ and $D_{N-1}$ and components $[M_1]_{1\alpha_1}^{i_1}$ and $[M_2]_{\alpha_{N-1}1}^{i_N}$, respectively). Given the above $N$ tensors, the MPS ansatz assumes that the $q^N$ wavefunction amplitudes in \eqref{eq:psi} can be written as
\begin{eqnarray} \label{eq:MPS1}
\psi^{i_1i_2\cdots i_N} &=& \sum_{\{\alpha\}} [M_1]_{1\alpha_1}^{i_1} [M_2]_{\alpha_1\alpha_2}^{i_2} \cdots [M_2]_{\alpha_{N-1}1}^{i_N} \\
&=& [M_1]^{i_1} \cdot [M_2]^{i_2} \cdot~.\,.\,.~\cdot [M_2]^{i_N}, \label{eq:MPS2}
\end{eqnarray}
where in the first line we have written explicitly both all the indices of each tensor and the sum over the matrix indices $\{\alpha\} = \alpha_1, \alpha_2, \cdots, \alpha_{N-1}$, whereas in the second line we regarded each $[M_{k}]^{i_k}$ as a matrix (except for $[M_{1}]^{i_1}$ and $[M_{N}]^{i_N}$, which are vectors) and used the matrix product symbol `$\cdot$' to represent matrix-matrix multiplication (respectively, matrix-vector multiplication). Notice that the name `matrix product state' of this variational ansatz comes from the fact that it expresses the wavefunction amplitudes as the matrix product \eqref{eq:MPS2}.

Intuitively, the MPS ansatz assumes that the state $\ket{\psi}$ has a restricted amount of entanglement. Indeed, the so-called bond dimension $D_k$ limits how much entanglement the ansatz can represent between two parts of the system, namely between a part containing the first $k$ sites (from site $1$ to site $k$) and another part containing the rest of sites (from site $k+1$ to site $N$). In particular, the larger the bond dimension $D_k$ is, the more entanglement the MPS can account for between these two parts, and thus the more capable the ansatz is to represent entangled many-body wavefunctions. On the other hand, tensor $M_{k}$ contains $q \times D_{k-1} \times D_k$ variational parameters, so that the cost of storing the MPS grows with the bond dimensions. Quite often, the bond dimension $D_k$ is chosen according to the rule
\begin{eqnarray}
D_k =  \left\{ \begin{array}{ll}
    q^k &  ~\mbox{for} ~~~1\, \leq k < k_1  \\
    D &   ~\mbox{for} ~~ k_1 \leq k \leq k_2 \\
    q^{N-k} &  ~\mbox{for} ~~ k_2 < k \leq N-1,
\end{array}  \right.
\end{eqnarray}
namely such that it grows exponentially with $k$ for small $k$ until it reaches some maximum allowed bond dimension $D$, and similarly with $N-k$ for large $k<N$. Given a choice of maximum bond dimension $D$, $k_1$ above is simply the smallest site index such that $D < q^{k_1}$, whereas $k_2$ is the largest site index such that $D < q^{N-k_2}$. In many applications the above prescription implies that most of the MPS tensors have size $q \times D^2$. Then the memory space required to store the MPS scales with the bond dimension $D$ and system size $N$ as $O(ND^2)$. 

The MPS ansatz comes with so-called gauge freedom \cite{cirac_matrix_2021,schollwock_density-matrix_2011, mcculloch_density-matrix_2007}, in that the amplitudes $\psi^{i_1i_2 \cdots i_N}$ are invariant under the simultaneous change $[M_k]^{i_k} \rightarrow [M_{k}]^{i_k} \cdot Q_k$ and $[M_{k+1}]^{i_{k+1}} \rightarrow Q_{k}^{-1} \cdot [M_{k+1}]^{i_{k+1}}$,
where $Q_k$ is any invertible $D_k \times D_k$ matrix and $Q^{-1}_k$ is its inverse. Indeed, the double replacement is easily seen to leave the matrix product $[M_k]^{i_k} \cdot [M_{k+1}]^{i_{k+1}}$ invariant in \eqref{eq:MPS2}, so that the wavefunction amplitudes are not changed. Using this gauge freedom, we can bring the MPS \eqref{eq:MPS} into the so-called {\it central} gauge \cite{cirac_matrix_2021,schollwock_density-matrix_2011} with respect to site $n$, 
\begin{equation} \label{eq:MPS_central}
    \{A_1,~ \cdots, A_{n-1}, C_{n}, B_{n+1},~ \cdots, B_{N}  \},
\end{equation}
where letters $A$ and $B$ are used to denote MPS tensors that satisfy the following orthogonality constraints:
\begin{eqnarray} \label{eq:left}
\sum_{i_k} \left([A_k]^{i_k} \right)^{\dagger} \cdot [A_k]^{i_k} &=& \unity ~~~~~\forall k < n,\\
  \sum _{i_k} [B_k]^{i_k} \cdot \left([B_k]^{i_k}\right)^{\dagger} &=& \unity ~~~~~\forall k > n.\label{eq:right}
\end{eqnarray}
The central tensor $C_n$ above does not satisfy any of the two relations. The central form is important in order to both simplify, and provide numerical stability to, the MPS optimization procedure, as briefly reviewed below.

The system's Hamiltonian $\mathcal{H}$ in \eqref{eq:H} can similarly be expressed in matrix product operator  (MPO) \cite{zwolak_mixed-state_2004, verstraete_matrix_2004, pirvu_matrix_2010, michel_schur_2010, hubig_generic_2017} form, given in terms of a sequence of $N$ order-4 tensors
\begin{equation}
    \{H_1, H_2, \cdots, ~H_N \},
\end{equation}
where each tensor $H_k$ has components $[H_k]^{mi_kj_k}_{m_{k-1}m_k}$. For fixed values of the physical indices $i_k$ and $j_k$, we can think of $[H_k]^{i_kj_k}$ as a $D'_{k-1} \times D'_{k}$ matrix. We again refer to $i_k$ and $j_k\}$ as \textit{physical} indices and to $m_{k-1}$ and $m_k$ as MPO \textit{bond} indices. The Hamiltonian coefficients in \eqref{eq:H} can then be written as
\begin{eqnarray}
\mathcal{H}^{i_1j_1i_2j_2\cdots i_Nj_N} &=& \sum_{\{m\}}[H_1]^{i_1j_1}_{1 m_1} [H_2]^{i_2j_2}_{m_1m_2}  \cdots [H_N]^{i_N j_N}_{m_{N-1}1} ~~~~~~~~~ \\
&=& [H_1]^{i_1j_1} \cdot [H_2]^{i_2j_2} \cdot ~.\,.\,.~ \cdot [H_N]^{i_N j_N}.  
\end{eqnarray}
Importantly, in this case the MPO representation is not used as a variational ansatz for the Hamiltonian $\mathcal{H}$, but as a convenient way of exactly representing it.

\subsection{Variational Energy Optimization}

The exact ground state $\ket{\psi_{\GS}}$ of Hamiltonian $\mathcal{H}$ is the state $\ket{\psi}$ that minimizes the expectation value of the energy, as given by $E(\ket{\psi}) \equiv \braket{\psi|\mathcal{H}|\psi} /\braket{\psi|\psi}$. Accordingly, an MPS approximation $\ket{\psi^\star_{\MPS}}$ to the ground state $\ket{\psi_{\GS}}$ is obtained by minimizing the energy
\begin{equation} \label{eq:E_MPS}
E(\ket{\psi_{\MPS}}) = \frac{\braket{\psi_{\MPS}| \mathcal{H}| \psi_{\MPS}}}{\braket{\psi_{\MPS} |\psi_{\MPS}}}    
\end{equation}
over the set of states $\ket{\psi_{\MPS}}$ that can be written as an MPS (for some fixed choice of bond dimensions $\{D_k\}$), 
\begin{equation} \label{eq:DMRG_goal}
    \ket{\psi^{\star}_{\MPS}} = \arg \min_{\ket{\psi_{\MPS}}} E(\ket{\psi_{\MPS}}).
\end{equation}
In the following we will outline the main steps of the DMRG algorithm, which aims at obtaining $\ket{\psi^{\star}_{\MPS}}$, and refer the reader to the literature \cite{white_density_1992, schollwock_density-matrix_2011} for more details. 

Starting from some initial state $\ket{\psi_{\MPS}}$ given by a set of MPS tensors $\{M_k\}$ in \eqref{eq:MPS}, the DMRG algorithm attempts to minimize the energy $E(\ket{\psi_{\MPS}})$ in \eqref{eq:E_MPS} by iteratively optimizing one MPS tensor at a time. More concretely, we first optimize the tensor on site $n=1$, then the tensor on site $n=2$, etc, all the way to the tensor on site $n=N$, in what is known as a \textit{forward sweep}. That is followed by a \textit{backward sweep}, progressing from the last site to the first one. For a given site $n$, the optimization proceeds as follows. First, the MPS is written in the central canonical form for that site, namely as in Eq. \eqref{eq:MPS_central}. Then the central tensor $C_n$ is replaced with a new tensor $C'_n$ that is chosen in a way as to optimally lower $E(\ket{\psi_{\MPS}})$ while keeping the rest of MPS tensors constant. This is accomplished by diagonalizing a linear operator using a Krylov-space method such as the Lanczos method (see Fig. \ref{fig:datadist} and text below for more details).

The DMRG algorithm proceeds with forward and backward sweeps until either the energy $E(\ket{\psi_{\MPS}})$ has converged to some desired accuracy, or a maximum sweep number has been reached. Although ideally one would like to obtain the optimal MPS ground state approximation $\ket{\psi^{\star}_{\MPS}}$ in Eq. \eqref{eq:DMRG_goal}, in practice it is understood that one must settle for a reasonably converged MPS ground state approximation, also denoted $\ket{\psi^{\star}_{\MPS}}$ in the rest of this paper.

\subsection{Computational Cost and Area Law}

For simplicity, from now on we assume a uniform MPS bond dimension $D$. Optimizing a central tensor $C_n \rightarrow C_n'$ (e.g. using Lanczos) as well as shifting the central canonical form \eqref{eq:MPS_central} from site $n$ to site $n+1$ (which can be accomplished using a polar, QR or singular value decomposition \cite{golub_singular_1970}) both have a computational cost $O(D^3)$, leading to a total cost per optimization sweep that scales with bond dimension $D$ and system size $N$ as $O(ND^3)$. It is important to emphasize the \textit{linear} scaling in $N$ at fixed $D$, as opposed to the \textit{exponential} scaling incurred in a brute-force ground-state computation using the $q^N$ amplitudes in \eqref{eq:psi}. However, in order to achieve some desired accuracy with an MPS computation, the bond dimension $D$ may need to be adjusted as a function of the system size $N$, that is, in general $D = D(N)$, in which case the scaling of computational resources may no longer be linear in $N$.

In order to gain further insight into computational costs, a useful rule of thumb is to think that the bond dimension $D$ must grow exponentially with the entanglement entropy $S$ \cite{vidal_entanglement_2003, peschel_calculation_2003, latorre_ground_2004} of the target ground state $\ket{\psi_{\GS}}$ for half of the system, that is $D \sim \exp(S)$. For a generic (non-local) Hamiltonian $\mathcal{H}$, the half-system entropy of the ground state is expected to grow linearly in the system size, $S \approx N$, scaling known as entanglement \textit{volume law}. In this case, the bond dimension must grow exponentially with system size, $D \sim \exp(N)$, and using an MPS representation is not computationally advantageous with respect to a brute-force computation. Luckily, ground states of local Hamiltonians often have a more forgiving scaling of entanglement entropy $S$ with system size $N$, known as the entanglement \textit{area law} \cite{bombelli_quantum_1986, srednicki_entropy_1993, amico_entanglement_2008,vidal_entanglement_2003, latorre_ground_2004, wolf_violation_2006,gioev_entanglement_2006, peschel_reduced_2009, eisert_colloquium:_2010} (sometimes with a logarithmic correction), which justifies the use of an MPS representation and the DMRG algorithm.

Specifically, in a $d$-dimensional cubic lattice made of $N = L^d$ sites, the half-system entropy of a state obeying the \textit{area law} scales as
\begin{equation} \label{eq:area_law}
    S \sim L^{d-1} = N^{\frac{d-1}{d}}~~~~~~~\mbox{(area law)}~~~~
\end{equation}
or, in the presence of a logarithmic correction, as 
\begin{equation} \label{eq:log_correction}
    S \sim  N^{\frac{d-1}{d}} \times \log N ~~~~\mbox{(log. correction)}.
\end{equation}
The above rule of thumb then indicates that the required MPS bond dimension should scale, respectively, as 
\begin{eqnarray}\label{eq:Dscaling1}
    D &\sim& \exp \left(N^{\frac{d-1}{d}}\right) ~~~~~~~~~~~~~~~~~~~~\mbox{(area law)}~~~~\\
    D &\sim& \exp \left(N^{\frac{d-1}{d}}\right) \times \mbox{poly}(N) ~~~~\mbox{(log. correction)}.~~~\label{eq:Dscaling2}
\end{eqnarray}

In $d=1$ dimensions, the ground state typically obeys an area law if the local Hamiltonian has a finite energy gap. In this case $S\sim N^0$ suggests that a constant bond dimension $D$, independent of the system size $N$, may suffice to accurately approximate the ground state $\ket{\psi_{\GS}}$ with an MPS, resulting in overall computational cost linear in $N$. For gapless Hamiltonians in $d=1$ dimensions, the ground state often exhibits a logarithmic correction to the area law, $S \sim \log(N)$, resulting in a polynomial bond dimension $D\sim $ poly$(N)$ and thus also a computational cost that grows as some power of the system size $N$. We conclude that DMRG is efficient for ground states of $d=1$ systems.

In $d=2,3$ dimensions, ground states of gapped and gapless local Hamiltonians often obey the entanglement area law, with some gapless systems (e.g. systems with a Fermi surface of dimension $d-1$) also display logarithmic corrections. For such systems, the above rule of thumb indicates that DMRG has cost that scales at least as $\exp(\sqrt{N})$ and $\exp(N^{2/3})$ in $d=2,3$ dimensions, respectively. Notice that, in spite of this exponential scaling of computational costs (in a fractional power of $N$), DMRG still has significant advantage with respect to the $\exp(N)$ scaling of a brute-force computation. 

While the area law (with possible logarithmic corrections in certain critical systems) has mostly been investigated in regularly structured lattice models, we expect it to also roughly apply to more generic systems, such as large molecules in $d=3$ dimensions, where DMRG is used in the context of quantum chemistry \cite{white_ab_1999, chan_density_2011, wouters_density_2014,szalay_tensor_2015, reiher_dmrg_nodate, olivares-amaya_ab-initio_2015, yanai_density_2015,chan_matrix_2016, white_multi-sliced_2019, baiardi_density_2020,brabec_massively_2020,barcza_dmrg_2020, qiu_hybrid_2021,goings_reliably_2022}.

In conclusion, DMRG is efficient in $d=1$ dimensions, where it is firmly established as the method of choice to compute ground states. On the other hand, DMRG scales exponentially as $\exp(\sqrt{N})$ and $\exp(N^{2/3})$ in $d=2,3$ dimensions. In spite of this unfavorable scaling, DMRG is actively used in restricted $d>1$ geometries such as thin two-dimensional strips and cylinders \cite{yan_spin-liquid_2011, stoudenmire_studying_2012, white_density_1998, white_hole_1997, white_energetics_1998, tohyama_stripe_1999, white_competition_1999, kampf_stripe_2001, white_phase_2000,weng_spin-liquid_2006, cincio_characterizing_2013, he_chiral_2014,simons_collaboration_on_the_many-electron_problem_solutions_2015} and small three-dimensional molecules \cite{chan_highly_2002, chan_exact_2003, legeza_qc-dmrg_2003, legeza_controlling_2003, chan_state---art_2004, moritz_convergence_2005, moritz_relativistic_2005, weser_chemical_2021, goings_reliably_2022}. In such cases, the massive computational cost of running DMRG constitutes the main roadblock to studying larger systems. As we show in this work using Tensor Processing Units, specialized hardware originally developed to accelerate and scale up machine learning workloads can be repurposed to also accelerate and scale up DMRG computations, significantly increasing the bond dimension $D$ that can be afforded. This allows us to use DMRG to more accurately address larger systems in $d>1$ dimensions.

\section{TPU cores, boards and pods}
\label{sec: tpus}

Tensor Processing Units (TPUs) are application specific integrated circuits (ASICs) developed by Google specifically for large scale machine learning applications \cite{TPUinfo, jouppi2017datacenter}. However, in recent times a growing number of papers have demonstrated their applicability to accelerating and scaling up other computationally intensive tasks, including large-scale dense linear algebra operations \cite{tpu_algebra}, the simulation of quantum circuits \cite{tpu_Z2field, tpu_circuit}, brute-force ground state computation and dynamics simulation in quantum many-body systems \cite{tpu_floquet, tpu_qphys, tpu_qhardware}, and quantum chemistry electronic structure computations using density functional theory \cite{tpu_qchem, tpu_matsci, tpu_proteins}. In this work we focus on TPUs of third generation, denoted v3 in the following. [After completion of our work, TPUs of fourth generation, with increased compute power, were made available. The results presented in this work can be straightforwardly generalized to TPU v4.] In the third generation, eight TPU v3 cores form a TPU board, and up to 256 TPU v3 boards can be connected into a TPU pod (with 2048 TPU v3 cores). 

A single TPU v3 core is equipped with two matrix-multiply units (MXUs) and 16 GB of on-chip, high-bandwidth memory (HBM). An MXU is a systolic array that can multiply matrices of size $128 \times 128$ natively, using multiplication of floating numbers in half precision (specifically, in \textit{brain float} 16 format, or bf16) and accumulation in single precision (fp32). Using 6 passes through the MXU, a single TPU core can however also deliver over 10 TFLOPS of single precision (fp32) matrix-matrix multiplication.

At the next level we find a TPU board, which is actually the smallest available configuration, with eight TPU v3 cores and one controlling host CPU machine. The eight cores are arranged into a 2d torus and, importantly, each core is connected to its neighbors through a fast Inter Core Interconnect (ICI) communication link (with 656GB/s bandwidth \cite{jouppi2017datacenter}). A TPU board has a total of 128 GB of HBM and can yield up to about 80 TFLOPS of single precision matrix-matrix multiplication \cite{tpu_algebra}.

Finally, up to 256 TPU boards (that is, up to 2048 TPU cores) can be joined into a TPU v3 pod, where the cores are again arranged on a 2d torus and directly connected to nearest neighbors with ICI links, with a total of 32 TB of HBM and near 20 PFLOPS of single precision matrix-matrix multiplication \cite{tpu_algebra}. One can also use a slice of a pod containing an intermediate number of TPU cores. For instance, in Fig. \ref{fig:measuredrt} we provide performance results and estimates for slices with 32, 128, 512 and 2048 cores. The largest TPU configuration we used in this work consisted in half a pod (that is, 1024 cores).

TPUs can be programmed using XLA \cite{XLA}, an optimized graph compiler that translates from roughly C-like commands called HLOs to roughly assembly-like equivalents called LLOs. The HLOs themselves may be written directly, but are usually instead ``traced" from any of several higher-level languages. For the DMRG work presented in this paper, we wrote the code with Jax \cite{Jax}, a NumPy like interface to XLA, following the single instruction multiple data (SIMD) paradigm.

\section{DMRG on TPUs}
\label{sec:dmrg on tpus}

The performance of our large-scale implementation of DMRG on multi-core TPU configurations is based on three main points: (i) individual MPS tensors (and other auxiliary tensors) are distributed through the available TPU cores; (ii) an out-of-core approach is adopted in order to more efficiently use the 16 GB of high bandwidth memory on each TPU core; (iii) tensor contractions are accelerated through parallelization.
\\

\subsection{Data distribution}

The largest data objects in a DMRG simulation are (a) the $N$ order-3 MPS tensors $M_n$ that contain the variational parameters of the ansatz and (b) two sets of $N$ auxiliary tensors $L_n$ and $R_n$, called left and right environment Hamiltonian tensors \cite{white_density_1992, schollwock_density-matrix_2011}, which given a choice of central site $n$, represent a contraction of MPS and MPO tensors for all sites $k<n$ and for all sites $k>n$, respectively. In components, we can write $[M_n]^{i}_{\alpha \beta}$, $[L_n]^m_{\alpha \beta}$, $[R_n]^m_{\alpha, \beta}$, where Greek letters are used to denote ``large'' indices, such as the MPS bond indices, with size $D$ that in our implementation could potentially be scaled up to $D \sim 10^5$, whereas Roman letters are used to denote ``small'' indices, namely the physical index $i$ taking $q$ values for $q=2$ in the examples below, and the MPO bond dimension $D'$, which in those examples grows up to $D'\sim 100$. 

\begin{figure}
  \includegraphics[width=0.4\columnwidth]{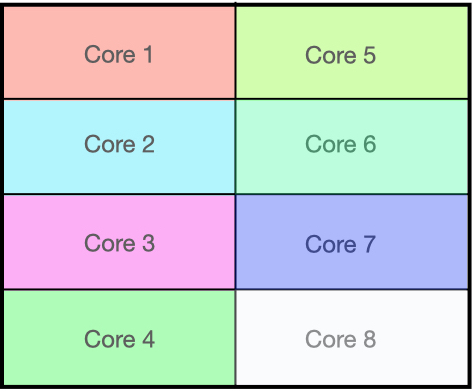}
  \;\;\;\;
   \includegraphics[width=0.4\columnwidth]{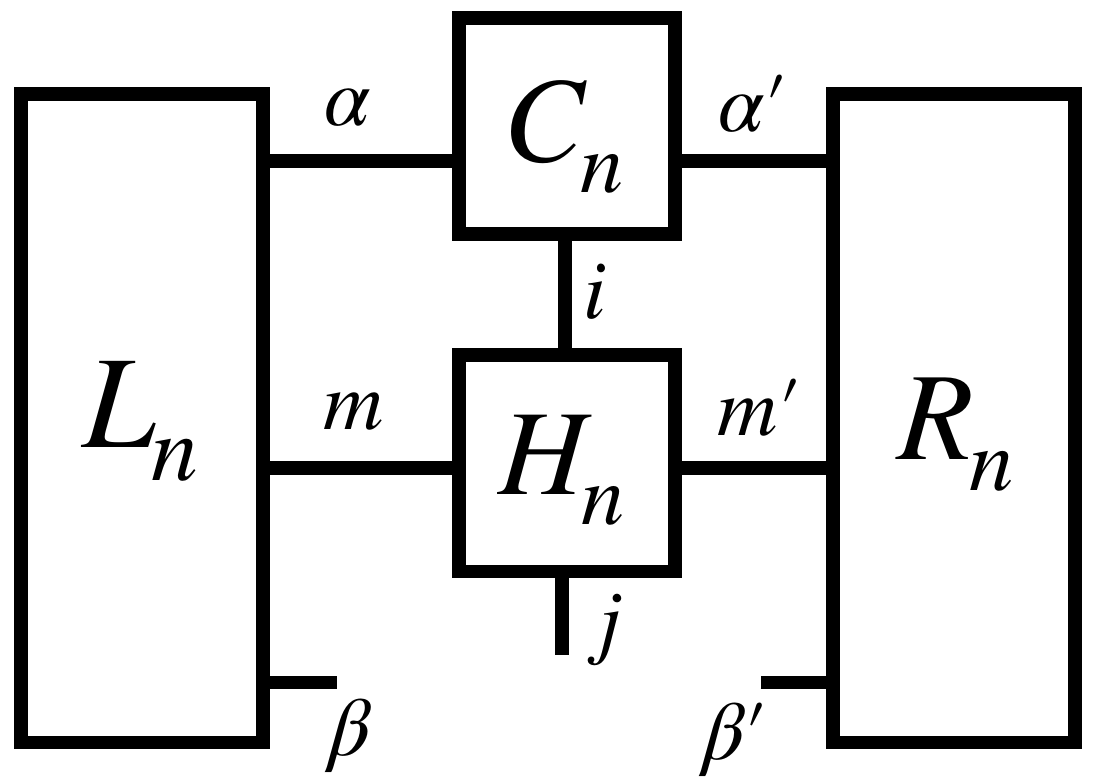}
  \caption{Left: Matrix distribution pattern on TPU cores. The figure shows the distribution pattern of a $D \times D$ matrix on 8 available TPU cores, arranged into a $4 \times 2$ torus. Each colored panel
  has shape $D/4 \times D/2$. Right: Tensor network diagram for a key tensor contraction in the DMRG algorithm, where one site MPS tensor (regarded as an effective vector wavefunction for a given site) is multiplied by three other tensors (corresponding to an effective Hamiltonian for that site).}\label{fig:datadist} 
\end{figure}

Let $T$ denote any of these order-3 tensors, with components $[T]^i_{\alpha\beta}$. In our implementation, we regard tensor $T$ as a collection $\{T_{i=1},T_{i=2}, \cdots \}$ of large matrices, where each matrix $T_i$ has components $[T_i]_{\alpha\beta}$ given by $[T]^{i}_{\alpha\beta}$. Each matrix $T_i$ is then distributed across all available TPU cores in a checkerboard fashion, as shown in 
\Fig{fig:datadist} for the case of eight TPU cores. Each matrix panel is stored in the high bandwidth memory of the corresponding TPU core. Since in SIMD code each matrix panel is expected to have the same size, matrix dimensions are chosen appropriately such that they can be evenly divided by the grid-shape of the TPU cluster. The motivation to distribute data in this way will become clear below, where tensor contractions are reduced to sequences of distributed matrix-matrix multiplications.

\subsection{Out-of-core approach}

The memory required to store the MPS tensors and the left and right environment Hamiltonian tensors scales as $O(ND^2q)$ and $O(ND^2D')$, respectively, where $N$ is the number of sites, $D$ the MPS bond dimension, $q$ the physical index dimension and $D'$ the MPO bond dimension. For large DMRG computations, these memory requirements can quickly become prohibitive, when compared to the total available HBM on TPUs. For example, for a system of two-level quantum degrees of freedom on each site (local dimension $q=2$) on a square lattice made of a $10\times 10$ grid (number of sites $N=100$) with 
a local Hamiltonian consisting only of a nearest neighbor fermionic hopping term 
(MPO bond dimension $D'=22$) and with an MPS bond dimension $D=2^{16} = 65,\!536$, the minimally required memory using single precision (4 bytes per fp32) grows to about $32$TB, which therefore already exhausts the maximally available 32TB HBM of an entire TPU v3 pod.

On the other hand, at any given time of a DMRG optimization sweep only a small number of such tensors are really required for processing on the TPUs. We have therefore adopted an out-of-core approach, where the bulk of the data is stored on hard drives (which are readily and cheaply available). For one optimization step of the DMRG algorithm, the necessary data is read from hard-drive into the HBM of the TPUs, where the relevant optimization step is executed, then the result is written back to the hard drive. To minimize the idling time of the TPUs, we utilize three simultaneous threads to perform DMRG optimization on the TPUs, and reading and writing of data from and to the disk.

\subsection{Distributed tensor contractions}

To illustrate how tensor contractions are performed in a distributed way, let us consider the contraction of the tensor network shown in the right panel of \Fig{fig:datadist}, which is the bottleneck operation in the DMRG algorithm. Given a site $n$, the tensor network contains the central MPS tensor $C_n$ for that site, as well as the left and right environment Hamiltonian tensors $L_n$ and $R_n$ and the MPO tensor $H_n$. Conceptually, we can think of the contraction of this tensor network as corresponding to a ``vector-matrix" multiplication, if we regard the central MPS tensor $C_n$ as representing a ``vector" (an effective wavefunction on site $n$) and the remaining three tensors as representing a ``matrix'' (an effective Hamiltonian $H_{\eff}$ on site $n$). This effective ``vector-matrix" multiplication needs to be performed several times as part of the Lanczos tridiagonalization (which aims to compute the ground state of the effective Hamiltonian $H_{\eff}$ on site $n$ as part of a single DMRG optimization step).

In order to proceed, we first preprocess the MPO tensor $H_n$, with components $[H_n]^{ij}_{\alpha \beta}$ which often vanish (sparse MPO), into a list of non-zero components $v = \{v_1, v_2, \dots \}$ and their corresponding multi-indices $p=\{(a_1,b_1,c_1,d_1), (a_2,b_2,c_2,d_2)\dots\}$ s.t. $M^{a_kb_k}_{c_kd_k}=v_k$

The tensor network contraction is then performed by looping over all elements in the lists,
see Algorithm \ref{alg:contraction}. Notice that for each non-zero value in $v$, we must perform two matrix-matrix multiplications involving matrices from tensors $L_n$, $C_n$ and $R_n$. As explained above, each of these matrices has been suitably distributed among the TPU cores. Then we multiply them using the \textit{scalable universal matrix multiplication algorithm} (SUMMA) algorithm  \cite{summa}, following a TPU implementation discussed in \cite{tpu_algebra}. Each iteration of the loop produces a different distributed matrix, which is weighted by the corresponding weight $v_k$ and added to one of the $q$ matrices that will constitute the final order-3 tensor with the result of the tensor network contraction.

This approach is particularly appealing for highly sparse MPO tensors, as one typically find when the MPO is encoding a local Hamiltonian of a lattice model. For dense MPO matrices, as e.g. appearing in some quantum chemistry applications of DMRG, a vectorized approach can be more efficient. 

Another important step in the DMRG algorithm is tensor orthogonalization \cite{schollwock_density-matrix_2011}, which is traditionally implemented using a QR decomposition or a singular value decomposition. In this work we chose to perform orthogonalization using instead a polar decomposition (which, in a so-called two-site DMRG approach can also be used for optimal tensor truncation). Further implementation details can be found in the Appendix.

\begin{figure}
\begin{algorithm}[H]
    \caption{Contraction algorithm}\label{alg:contraction}
    \begin{algorithmic}[1]
    \Function{contract($i,v,L, R, C$)}{}
    \State $M$ = zeros\_like($C$)\Comment{Container for storing the final contraction result}
    \For{$n$=0\dots\rm{len}($i$)-1} 
        \State $a,b,c,d = i[n]$
        \State $T$ = SUMMA(SUMMA($L[c,...],C[a,...]$), $R[d,...]$)
        \State $M[b,...]$+= $v[n]$*T \Comment{Accumulate matrix multiplications}
    \EndFor
    \State \Return $M$
    \EndFunction
    \end{algorithmic}
\end{algorithm}
\end{figure}

\section{Results}
\label{sec: results}

In order to benchmark our distributed implementation of DMRG on TPUs, we computed an MPS approximation $\ket{\psi^{\star}_{\MPS}}$ to the ground state $\ket{\Psi_{\GS}}$ of two different 2d square lattice models. The first one is a model of free spinless fermions on a lattice of size $10\times 10$, which can also be solved efficiently using the free fermion formalism, so that we have the exact solution to compare against. Its ground state displays a logarithmic correction to the area law, making this model extremely challenging from a DMRG perspective. The second model is the transverse field Ising model on a lattice of size $20 \times 20$, for which we do not have an exact solution, but other techniques can be used. The ground state of this model obeys an area law. This makes it less computationally demanding for DMRG, allowing us to consider a larger lattice.

The two models analyzed in this section are already well understood. We have chosen them mostly for two reasons. On the one hand, they are challenging from a DMRG perspective and, as such, can be used to meaningfully illustrate the use of very large bond dimension, as made available by TPUs. On the other hand, such models are often also used to benchmark other methods, including quantum monte carlo \cite{inglis_entanglement_2013} and numerical linked-cluster expansions \cite{kallin_entanglement_2013} or other tensor network algorithms such as those based on a tree tensor network (TTN) \cite{milsted_tensornetwork_2019}, the multi-scale entanglement renormalization ansatz (MERA) \cite{cincio_multiscale_2008,evenbly_entanglement_2009} and projected entangled pair states (PEPS) \cite{lubasch_algorithms_2014}. Benchmarking DMRG on the same models (although for different system sizes) enables useful comparisons.

\subsection{Free fermion model}

We first consider, on a $10\times 10$ square lattice with $N=100$ sites, the nearest neighbor Hamiltonian 
\begin{equation}
\mathcal{H}_{\SF} = -\sum_{\langle i, j\rangle} \hat c_i^{\dagger}\hat c_j+\mu \sum_i \hat c_i^{\dagger} \hat c_i \label{eq:SF}
\end{equation}
with $\hat c_i$  (anti-commuting) fermionic annihilation operators and $\mu$ the chemical potential. This model describes a system of non-interacting/free electrons that can hop from each site to its nearest neighboring ones, where the value of $\mu$ can be tuned to determine the number of electrons in the ground state (e.g. $N/2=50$ particles for $\mu = 0$). Using the free fermion formalism, the quadratic Hamiltonian \eqref{eq:SF} can be numerically diagonalized with computational cost that scales just as $O(N^3)$, instead of the generic $O(\exp(N))$ of a brute-force diagonalization. This is in contrast with the interacting case (e.g. if we added quartic terms to the above Hamiltonian), where the free fermion formalism can no longer be used. Here, the ground-state energy from the $O(N^3)$ diagonalization will be used to assess the accuracy of the DMRG result. 

It is important to emphasize that, despite the lack of interactions, computing the ground-state of Hamiltonian \eqref{eq:SF} is still a formidable challenge from the perspective of the DMRG algorithm. Indeed, for sufficiently small value of $|\mu|$ this Hamiltonian is seen to describe a system with a one-dimensional Fermi surface, which results in the presence of a large number of gapless excitations. As such, its ground state $\ket{\psi_{\GS}}$ displays a logarithmic correction to the area law \cite{vidal_entanglement_2003, wolf_violation_2006}, implying that an accurate MPS approximation $\ket{\Psi^{\star}_{\MPS}}$ requires a bond dimension $D$ expected to scale faster than $O(\exp(\sqrt{N}))$, see Eqs. \eqref{eq:log_correction} and \eqref{eq:Dscaling2} for $d=2$. This is the strongest scaling of ground state entanglement (and bond dimension $D$) observed to naturally occur in condensed matter systems in $d=2$ dimensions. Thus, as far as DMRG is concerned, this non-interacting/free lattice model is not easier than a strongly interacting/strongly correlated lattice model.

\begin{figure}
  \includegraphics[width=1\columnwidth]{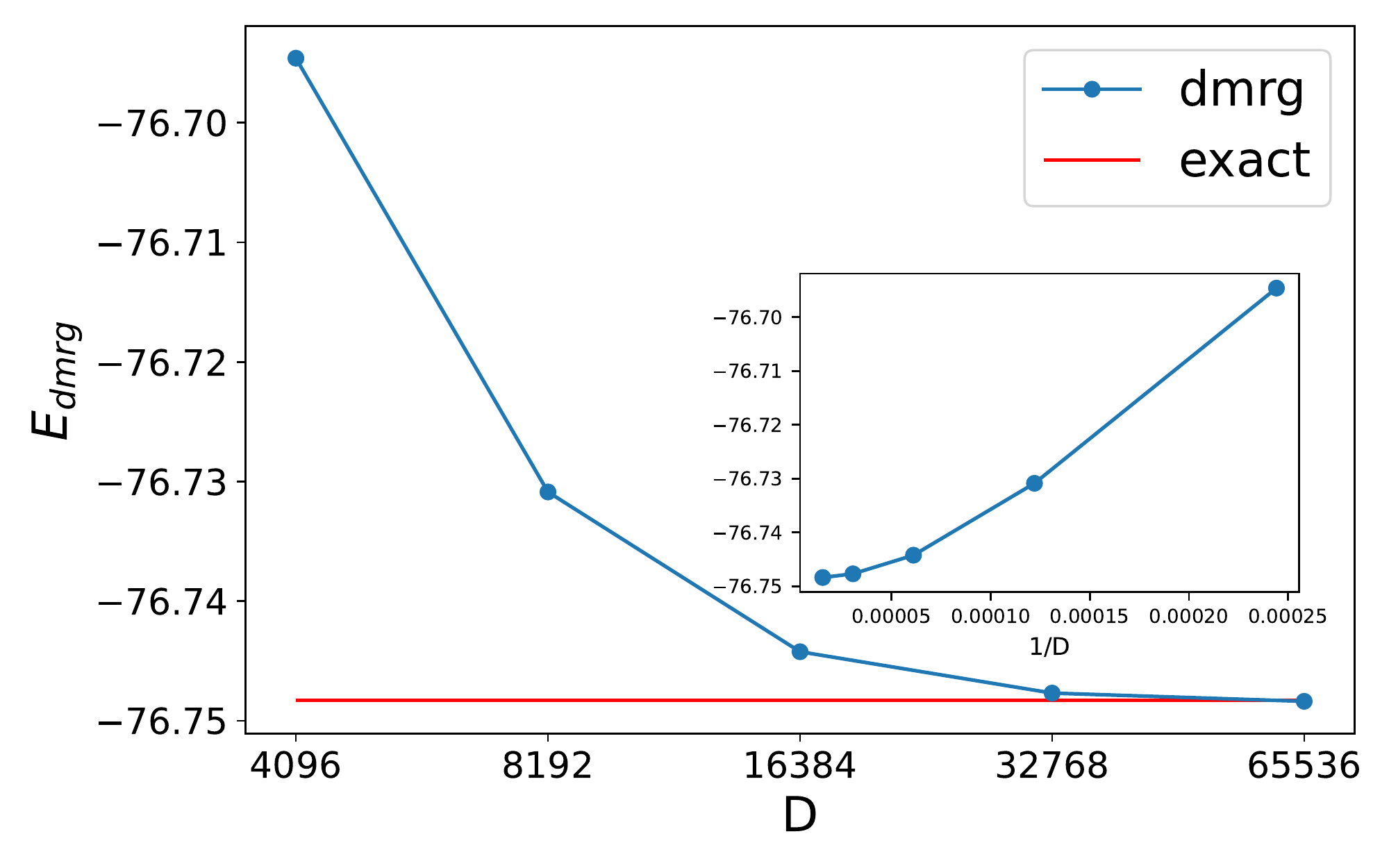}
  \includegraphics[width=1\columnwidth]{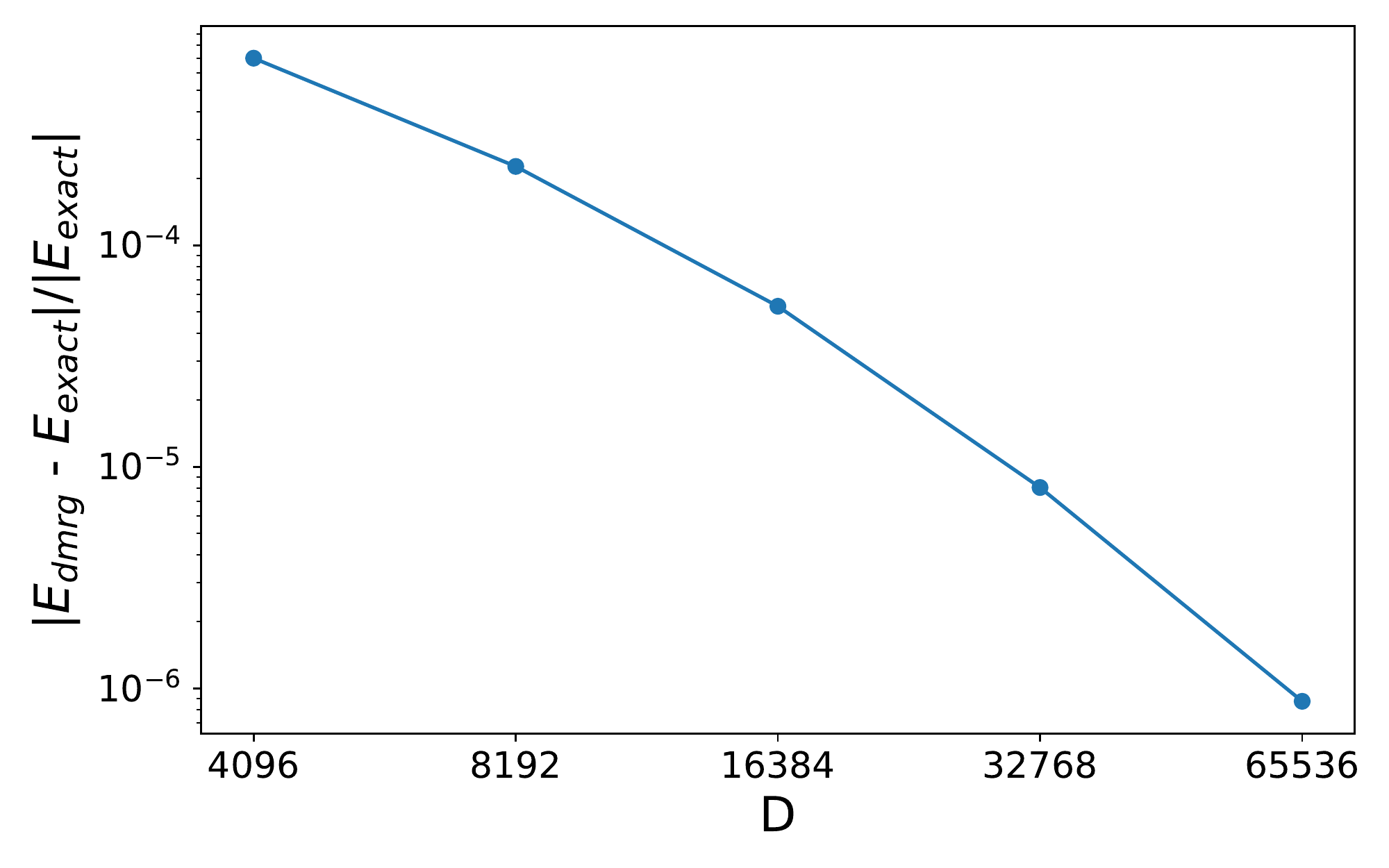}
  \caption{Top: Convergence of the DMRG ground state energy towards the exact value as a function of bond dimension $D$ for a 2d system of spinless free fermions on a 10 $\times$ 10 square lattice. Bottom: Convergence of the relative error in energy for the same system.}\label{fig:sfenergies}
\end{figure}

\Fig{fig:sfenergies} presents the DMRG approximations $E(\ket{\psi^{\star}_{\MPS}})$ for the ground-state energy $E(\ket{\psi_{\GS}})$ for $\mu=0$ (half filling). The Hamiltonian was encoded in an MPO with bond dimension $D'=22$. The top panel shows the converged DMRG energy as a function of the bond dimension $D$. The red line denotes the numerically exact value obtained from a $O(N^3)$ diagonalization. The bottom panel in \Fig{fig:sfenergies} shows the evolution of the relative error of this energy as a function of bond dimension $D$. At $D=2^{16}$ the approximation achieves a relative accuracy of less than one part in a million. Note that simulations were carried out in single precision, limiting the maximum achievable accuracy to about $10^{-7}$. To the best of our knowledge, these are the largest DMRG computations (in terms of bond-dimension) to date. Results for $D=2^{16}=65,\!536$ were obtained within roughly 23 hours on a slice of a TPU v3 pod made of $1,\!024$ cores. We used seven sweeps and a Krylov dimension of 10 for the sparse diagonalization required for optimizing each tensor.

A remark regarding internal symmetries in the MPS representation is in order. Hamiltonian $\mathcal{H}_{\SF}$ commutes with the particle number operator $\mathcal{N} = \sum_i \hat c_i^{\dagger} \hat c_i$, indicating particle number preservation, an internal U$(1)$ symmetry generated by $\mathcal{N}$. Thus its ground state $\ket{\psi_{\GS}}$ also has a well-defined particle number $N_{\GS}$, $\mathcal{N}\ket{\psi_{\GS}} = N_{\GS} \ket{\psi_{\GS}}$. This can be exploited in DMRG \cite{mcculloch_density-matrix_2007, singh_simulation_2010, singh_tensor_2010, singh_tensor_2012, schmoll_programming_2020}. Indeed, by specializing the MPS tensors $M_k$ to be themselves invariant/co-variant under the Abelian U$(1)$ symmetry group, we can ensure that the MPS representation is exactly symmetric with the correct number $N_{\GS}$ of particles, $\mathcal{N}\ket{\psi_{\MPS}} = N_{\GS} \ket{\psi_{\MPS}}$. In addition, this confers each MPS tensor a block-sparse structure that significantly reduces the number of variational parameters to be optimized, as well as the required computational cost. 

Our current distributed implementation of DMRG on TPUS does not enforce or exploit the above model's internal U$(1)$ symmetry. Our goal here is to benchmark the performance of DMRG in a way that the results are representative of a more general 2d lattice model, where such internal symmetry may not be present (see e.g. our next example). We foresee nevertheless no obstruction to incorporating particle conservation in our current implementation.

\subsection{Transverse field Ising model}

\begin{figure}
   \includegraphics[width=1\columnwidth]{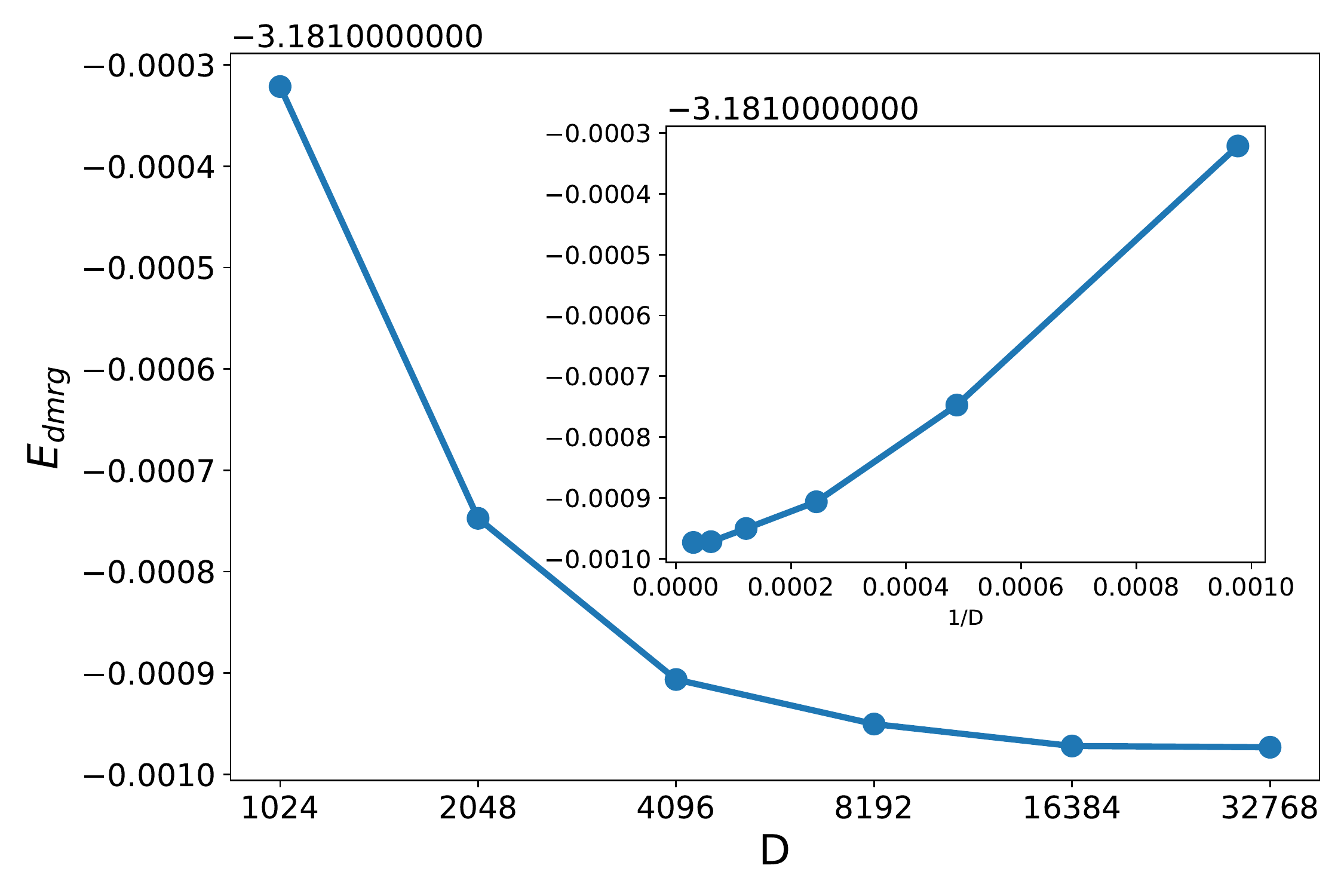}
  \caption{Convergence of the DMRG ground state energy per site as a function of bond dimension $D$ for the critical transverse field Ising model on a $20 \times 20$ lattice, leading to an energy density of $-3.18197(2)$. For comparison, Ref. 
  \cite{lubasch_algorithms_2014} obtained the values -3.17210(1) and -3.18243(1) for the energy density using a PEPS simulation on an $11 \times 11$ and $21 \times 21$ lattices, respectively.
  }\label{fig:isingenergies}
\end{figure}

\begin{figure}
  \includegraphics[width=1\columnwidth]{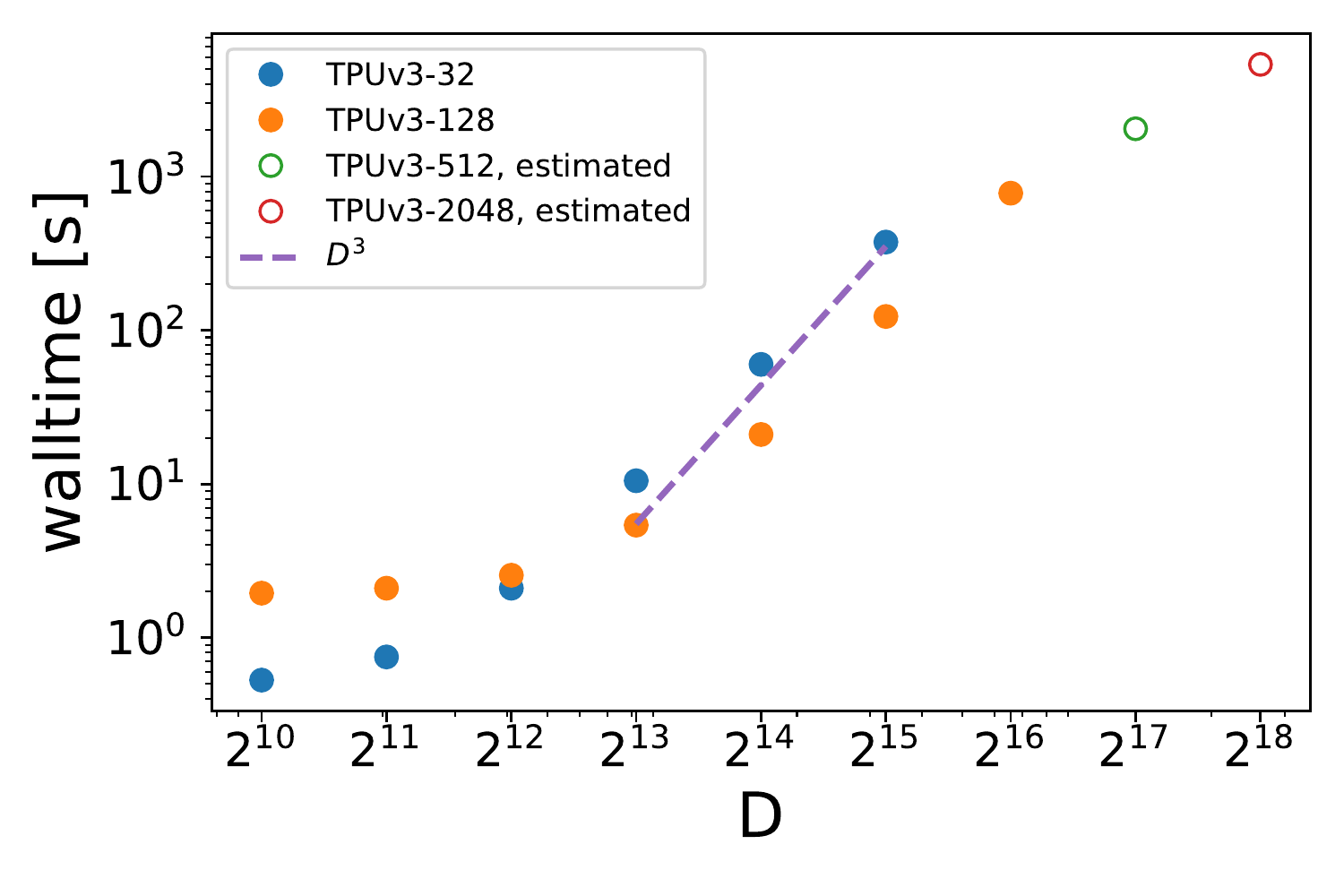}
  \caption{Measured and estimated runtimes per DMRG optimization step (including Lanczos update, tensor orthogonalization
    and update of the effective environment Hamiltonian) on different TPU cluster sizes. The solid line is proportional to the expected $D^3$ scaling
    of the DMRG algorithm.}\label{fig:measuredrt}
\end{figure}

As a second benchmark, we have also considered the transverse field Ising model,
\begin{equation}
\mathcal{H}_{\TFI} = -\sum_{\langle i, j\rangle} \hat\sigma^z_i\hat \sigma^z_j  + B \sum_i \hat \sigma_i^x\label{eq:tfi}
\end{equation}
on a $20\times 20$ square lattice with $N=400$ sites. Here $\hat \sigma^{x}_i$ and $\hat \sigma^{z}_i$ are Pauli matrices and $B$ the magnetic field strength. This model represents a system of spin-$\frac{1}{2}$ quantum spin degrees of freedom with ferromagnetic interaction $\hat\sigma^z_i\hat \sigma^z_j$ between nearest neighbor spins and subject to an external transverse magnetic field. This model is invariant under spin-flip transformations (internal Z$_2$ symmetry) generated by the unitary operator $U = \prod_i \hat \sigma_i^x$, which we again do not enforce or exploit in our DMRG implementation. The model has a quantum critical point for $B_c \approx 3.04$, and thus we expect the ground state $\ket{\psi_{\GS}}$ at or near this value of the magnetic field to be robustly entangled. Since there is no Fermi surface, the ground state entanglement entropy scales as an area law without logarithmic corrections, as previously confirmed using other methods \cite{cincio_multiscale_2008, evenbly_entanglement_2009,inglis_entanglement_2013,kallin_entanglement_2013}. Accordingly, the bond dimension $D$ required for an accurate approximation $\ket{\psi^{\star}_{\MPS}}$ scale as $O(\exp(\sqrt{N}))$, see Eqs. \eqref{eq:area_law} and \eqref{eq:Dscaling1} for $d=2$. This is still a very challenging computations for DMRG, but the milder scaling of the entanglement entropy (compared to the free fermion model above) allows us to consider a larger lattice.

\Fig{fig:isingenergies} shows the DMRG approximation $E(\ket{\psi^{\star}_{\MPS}})$ for the (unknown) ground state energy $E(\ket{\psi_{\GS}})$ of the transverse field Ising Hamiltonian \eqref{eq:tfi} on a 20 by 20 square lattice, at a near-critical magnetic field strength of $B=3.0$ and exactly encoded in an MPO with bond dimension $D'=22$. The plot shows how the converged DMRG energy per site appears to saturate to a constant as we increase the bond dimension $D$, in clear analogy with Fig. \ref{fig:sfenergies} for the spinless fermion model, where such saturation was to the correct value of the ground state energy. While this model cannot be solved analytically, results from numerical studies using e.g. Monte-Carlo or tensor network methods \cite{lubasch_algorithms_2014} are available, albeit not at the exact same system size. At bond dimension $D=2^{15}=32,\!768$ our simulations already reached the maximum level of achievable accuracy within single precision arithmetic. 

\subsection{Scaling of runtimes}

Finally, \Fig{fig:measuredrt} shows measured and estimated runtimes, for fixed MPO bond dimension $D'=22$ and as a function of the MPS bond dimension $D$, for the update of a single MPS tensor. These times include the time to perform the Lanczos tridiagonalization, the orthogonalization of the optimized tensor, and the update of one left or right block Hamiltonian. 

At fixed TPU configuration (fixed color in the plot), the runtimes are seen to scale with the MPS bond dimension $D$ as $D^3$. On the other hand, if every time that we double the bond dimension we quadruple the number of TPU cores (see data points for $D = 2^{15}, 2^{16}, 2^{17}$ and $2^{18}$ for 32, 128, 512 and 2048 cores, respectively), then the runtimes grow approximately only linearly in $D$. The type of scaling is key to reaching unprecedentedly large bond dimensions with affordable run times.

\section{Summary and discussion}
\label{sec: discussion}

We have presented an implementation of the DMRG algorithm on Google's Tensor Processing Units (TPUs). Our implementation leverages the distributed accelerated hardware
and high-bandwidth memory of a TPU cluster to perform DMRG simulations at unprecedented scale, speed and accuracy. We benchmarked the implementation on two problems that are notoriously difficult from a DMRG perspective, namely a system of spinless fermions on a $10 \times 10$ square lattice known to display a logarithmic correction to the area law, and the (near-) critical transverse field Ising model on a $20 \times 20$ lattice. We performed simulations with bond dimensions of up to $D=2^{16}$, to the best of our knowledge the largest ever simulated bond dimension so far. We obtained converged results at these bond dimensions in just less than a day. We estimate that such simulations would take months on standard shared-memory hardware with up to a few dozens of CPU cores, using highly efficient, compiled code. Our results show that compute clusters of hardware accelerators can be leveraged very efficiently for tensor network computations. For our demonstration we used TPUs, but we would like to emphasize that similar results can be obtained with a cluster of tightly connected Graphic Processing Units (GPUs).

There are several obvious ways to further improve the performance of our implementation of DMRG on TPUs. On the one hand, at an algorithmic level, we already mentioned that one can exploit the internal symmetries of a model (e.g. internal U(1) symmetry corresponding to particle number conservation in the 2d free fermion model \eqref{eq:SF}). Incorporating the internal U(1) symmetry into our implementation of DMRG will lead to a substantial reduction of variational parameters and run times for the same (effective) bond dimension. At fixed TPU configuration, we expect to then be able to further increase the maximal bond dimension $D$ by perhaps up to $\sim5-10\times$. On the other hand, the largest TPU configuration used in this work was made of 1024 cores, or half a TPU v3 pod. Using a full TPU v3 pod with 2048 cores would result in a roughly $2\times$ faster computation at fixed bond dimension $D$. Alternatively, it would allow us to increase the largest $D$ by a factor $\sqrt{2}$. While conducting our simulations, Google announced the fourth generation of TPUs, which are currently available. A TPU v4 pod (with 8192 TPU v4 cores) would allow for an additional $2\times$ increase of the maximal bond dimension $D$ at comparable runtimes. On the other hand, a superpod of NVIDIA's DGX nodes (with each DGX node containing eight A100 or H100 GPUs) could be utilized in a similar way to reach even larger bond dimensions.

It is worth pointing out that MPS algorithms similar to DMRG form also the basis for more sophisticated tensor networks approaches like e.g. projected entangled pair states for 2d quantum lattice systems \cite{verstraete_renormalization_2004}, and the availability of fast, large-scale MPS algorithms hence directly impacts not only DMRG but the field of tensor network algorithms as a whole. 

In conclusion, the large-scale implementation of DMRG with TPUs presented in this paper may have profound impacts in fields such as condensed matter physics \cite{yan_spin-liquid_2011, simons_collaboration_on_the_many-electron_problem_solutions_2015}, quantum chemistry \cite{white_ab_1999, chan_density_2011, wouters_density_2014,szalay_tensor_2015, reiher_dmrg_nodate, olivares-amaya_ab-initio_2015, yanai_density_2015,chan_matrix_2016, white_multi-sliced_2019, baiardi_density_2020,brabec_massively_2020,barcza_dmrg_2020, qiu_hybrid_2021,goings_reliably_2022} and material science \cite{ganahl_efficient_2015,ganahl_chebyshev_2014, bauernfeind_fork_2017} to machine learning \cite{gao_quantum_nodate, cheng_tree_2019, stoudenmire_supervised_2017, stoudenmire_learning_2018, glasser_probabilistic_2019, vieijra_generative_2022, roberts_tensornetwork_2019, han_unsupervised_2018}, where MPS and tensor network algorithms are either well established or rapidly gaining traction.
\\
\section{Acknowledgments}
The authors would like to thank {\"O}rs Legeza and Steve White for useful discussions and comments.
The authors also want to thank the JAX team for their continuous support, in particular Skye Wanderman-Milne and Jake Bradbury.
This research was supported with Cloud TPUs from Google’s TPU Research Cloud (TRC). Sandbox@Alphabet was a team within the Alphabet family of companies, which includes Google, Verily, Waymo, X, and others. 
G.V. is a CIFAR fellow in the Quantum Information Science Program, a Distinguished Invited Professor at the Institute of Photonic Sciences (ICFO), and a Distinguished Visiting Research Chair at
Perimeter Institute. Research at Perimeter Institute is supported by the Government of Canada through the Department of Innovation, Science and Economic Development and by the Province of Ontario through the Ministry of Research, Innovation and Science. 

\bibliography{references}

\clearpage
\newpage
\appendix
\section{Tensor orthogonalization}
\label{sec: ortho}

A crucial step in the DMRG method is the \textit{orthogonalization} of an optimized MPS tensor $M$ with components $M^{i}_{\alpha\beta}$. That refers to either one of the following decompositions:
\begin{eqnarray}
[M]^{i}_{\alpha \beta} &=& \sum_{\gamma} [A]^{i}_{\alpha \gamma} [R^{(l)}]_{\gamma \beta} ~~~  (\mbox{left orth.}) \label{eq:left2} \\
{[M]}^{i}_{\alpha\beta} &=& \sum_{\gamma} [R^{(r)}]_{\alpha \gamma} [B]^{i}_{\gamma\beta} ~~~  (\mbox{right orth.}) \label{eq:right2}    
\end{eqnarray}
where tensors $A$ and $B$ satisfy the left and right isometric constraint in \Eq{eq:left} and \Eq{eq:right}, respectively, which we rewrite here in components as
\begin{eqnarray} \label{eq:left3}
\sum_{i,\gamma} \left([A]^{i}_{\gamma \alpha}\right)^* [A]^{i}_{\gamma \beta} &=& \delta_{\alpha\beta}, ~~~ (\mbox{left isometry})\\
  \sum _{i, \gamma} [B]^{i}_{\alpha \gamma}  \left([B]^{i}_{\beta\gamma}\right)^{*} &=& \delta_{\alpha\beta}. ~~~ (\mbox{right isometry}) \label{eq:right3}
\end{eqnarray}
In the following we will focus on the left orthogonalization, Eqs. \eqref{eq:left2} and \eqref{eq:left3}. The right orthogonalization in Eqs. \eqref{eq:right2} and \eqref{eq:right3} can be obtained similarly. For the rest of this Appendix we regard tensor $M$ as a matrix, defined to have coefficients
\begin{equation} \label{eq: reshaping}
[M]_{(i\alpha)\beta} = [M]^{i}_{\alpha \beta}, 
\end{equation}
where we joined tensor indices $i$ and $\alpha$ into a single matrix index $(i\alpha)$ (notice that prior to that joining $i$ and $\alpha$, index $i$ was changed from an upper index to a lower index, which is a trivial change in our setting). This corresponds to a so-called \textit{reshaping} operation, which turns tensor $M$ into a matrix (also denoted $M$!) according to Eq. \eqref{eq: reshaping}. Notice that matrix $M$ is in general rectangular. For instance, at constant MPS bond dimension $D$, it has shape $(q D) \times D$, where $q$ is the dimension of the vector space describing one site of the lattice, so that it has $q\times$ more rows than columns. Our goal is to obtain $D$ orthonormal columns, in the sense of Eq. \eqref{eq:left3}.

Two popular approaches to orthonormalize $M$ are the use of either a QR decomposition or a singular value decomposition (SVD) of $M$. A TPU distributed version of QR or SVD was not available to us at the time we implemented DMRG. We were able to implement instead a TPU distributed version of the polar decomposition, which requires mostly distributed matrix-matrix multiplications and additions. The polar decomposition of $M$ re-expresses this matrix as a product of an isometric $(qD) \times D$ matrix $U$ and a positive semi-definite hermitian $D \times D$ matrix $H$, i.e.
\begin{equation}
  M = U \cdot H. \label{eq:polar}
\end{equation}
We can then obtain tensor $A$ and matrix $R^{(l)}$ in Eqs. \eqref{eq:left2} and \eqref{eq:left3} from $U$ and $H$ simply according to
\begin{eqnarray}
[A]^{i}_{\alpha \beta} = [U]_{(i\alpha) \beta},~~~~~~[R^{(l)}]_{\alpha \beta} = [H]_{\alpha \beta}.
\end{eqnarray}

The polar decomposition can be obtained by first normalizing $M$ into $X_0$ so that its largest singular value is upper-bounded by 1, namely 
\begin{equation}
    X_0 = M/z, ~~~z = ||M|| = \sqrt{\mbox{tr} \left( M^{\dagger} \cdot M \right) },
\end{equation}
and then converging the {\it Newton-Schultz} iteration:
\begin{align}
    X_{i+1} = X_i \cdot \left(\frac{3}{2}\mathbbm{1} - \frac{1}{2} X^{\dagger}_i \cdot X_i\right).\label{eq:newtonschultz}
\end{align}
It is easily verified that each iteration step \Eq{eq:newtonschultz} applies the polynomial
$P(x) = \frac{3}{2}x - \frac{1}{2} x^3$ to the (renormalized) singular values of $M$, while preserving its left and right singular vectors. Iterative application $\lambda_{i+1} = P(\lambda_i)$ maps initial real numbers $\lambda_0$ in the interval $(0,1]$ to 1, in the limit $i\rightarrow \infty$ (when a singular value $x$ is not too small, the iteration will turn it into 1 \textit{quadratically}, that is with a deviation from 1 that is suppressed quadratically in the number of iterations). \Eq{eq:newtonschultz} hence converges to the polar factor $U$ of $X_0$ (and thus of $M$) \cite{NakatsukasaHighamStable}. We can then also obtain $H$ from $M$ and $U$ simply using $H = U^{\dagger}\cdot M$.

It is instructive to relate the polar decomposition to the SVD of $M$, given by $M = W\cdot S \cdot V$, where $S$ is a diagonal matrix with the singular values and $W$ and $V$ are unitary (or isometric) matrices. The unitary (or isometric) and Hermitian factors $U$ and $H$ then read
\begin{equation}
    U = W\cdot V,~~~H = V^{\dagger}\cdot S \cdot V,
\end{equation}
where we used that
\begin{equation}
    M = W\cdot S \cdot V = (W\cdot V) \cdot  (V^{\dagger}\cdot S \cdot V) = U \cdot H.\nonumber
\end{equation}
We note that the case of singular values which are identically zero can be approximately addressed by adding to $M$ a diagonal constant perturbation of magnitude equal to machine precision $\epsilon$. 
 
Iteration \Eq{eq:newtonschultz} requires only matrix multiplications, transpositions, matrix addition and complex conjugation as fundamental operations. In the distributed setting, we implement the first two using the well-known SUMMA algorithm \cite{summa} for distributed matrix multiplications (SUMMA can also handle the case of multiplication of transposed matrices). Matrix addition and matrix complex conjugation of distributed matrices is trivial in that it can be carried out locally on each core.
\begin{figure}
\begin{algorithm}[H]
    \caption{Newton-Schultz iteration for the polar factor of $M$}\label{alg:polar}
    \begin{algorithmic}[1]
    \Function{polar\_factor($M$)}{}
    \State $z = \lVert M\rVert$
    \State $M\leftarrow M/z$ 
    \State $q=M$.shape[0]
    \State converged = False
    \While{not converged}
        \State $T$= zeros\_like($M[0,...]$)
        \For{$i$=0\dots $q-1$}
            \State $T$+=SUMMA($M[i,...]$,$M[i,...]$, 
            \State \quad\quad\qquad\qquad \; herm\_A=True, \Comment{Hermitian transpose of $A$}
            \State \quad\quad\qquad\qquad \;\;herm\_B=False)
        \EndFor
        \For{i=0\dots $q-1$}
            \State $M$[i,...] $\leftarrow\frac{3}{2}M$[i,...] -$\frac{1}{2}$SUMMA($M$[i,...], $T$)
        \EndFor
        \State converged = CHECK\_UNITARITY($M$)
    \EndWhile
    \State \Return $M$
    \EndFunction
    \end{algorithmic}
\end{algorithm}
\end{figure}

\section{Tensor truncation}
\label{sec: truncation}

Another operation of central importance in some implementations of DMRG (and, more generally, in many other tensor networks algorithms) is {\it truncation} of a matrix, namely rank reduction by retaining only its largest singular values. In DMRG, this is needed in the context of a two-site update. The DMRG implementation described in this paper corresponds to a one-site update and does not require tensor truncations, but here we explain how to implement them for completeness. 

Consider a matrix $M$ and its singular value decomposition (SVD),
\begin{equation} \label{eq: svd}
    M = W \cdot S \cdot V, 
\end{equation}
where $W$ and $V$ are unitary (or isometric) matrices and $S$ is a diagonal matrix with the singular values $s_{\alpha}$ of $M$ in its diagonal, that is $[S]_{\alpha\alpha} = s_{\alpha}$, organized in decreasing order $s_1 \geq s_2 \geq \cdots \geq s_m \geq 0$. In components, the SVD reads
\begin{equation} \label{eq: svd2}
    [M]_{\alpha \beta} = \sum_{\gamma=1}^{m} [W]_{\alpha \gamma} ~s_{\gamma}~ [V]_{\gamma \beta}, 
\end{equation}

Before proceeding further, we note here that in tensor network algorithms, the matrix $M$ to be truncated will often result from reshaping a higher-order tensor (e.g. an order-4 tensor assigned to two adjacent sites of a lattice in a two-site DMRG update). However, the specific origin of $M$ is not important in our discussion below. 

A $\delta$-truncated singular value decomposition of $M$ corresponds to the SVD of another matrix $\tilde{M}$ obtained by keeping only the singular values $s_{\alpha}$ of $M$ that are larger than $\delta$. Suppose that there are $m'$ (with $m' \leq m$) such singular values. Then $\tilde{M}$ is defined through
\begin{equation} \label{eq: truncated svd}
    [\tilde{M}]_{\alpha \beta} = \sum_{\gamma=1}^{m'} [\tilde{W}]_{\alpha \gamma} ~s_{\gamma}~ [\tilde{V}]_{\gamma \beta}, 
\end{equation}
where $\tilde{W}$ and $\tilde{V}$ are obtained from $W$ and $V$ by keeping only their first $m'$ columns and rows, respectively. That is,
\begin{eqnarray}
 \left[\tilde{W}\right]_{\alpha \gamma} &=& [W]_{\alpha \gamma} ~~~\mbox{for all $\alpha$, } ~ 1 \leq \gamma \leq m',\\
 \left[\tilde{V}\right]_{\gamma \beta} &=& [V]_{\gamma \beta} ~~~~\mbox{for all $\beta$, } ~ 1 \leq \gamma \leq m'.
\end{eqnarray}
We can similarly define a truncated singular value matrix $\tilde{S}$, of size $m'\times m'$, as the diagonal matrix that contains the $m'$ singular values $s_{\alpha}$ organized in decreasing order, such that
\begin{equation}
    [\tilde{S}]_{\gamma\gamma} = [S]_{\gamma\gamma} = s_{\gamma},~~~1\leq \gamma \leq m'. 
\end{equation}
Matrix $\tilde{M} = \tilde{W} \cdot \tilde{S} \cdot \tilde{V}$ can then be seen to be the rank-$m'$ best approximation to $M$, in that the difference matrix $\Delta \equiv M-\tilde{M}$ has the smallest possible norm $||\Delta|| = \sqrt{ \mbox{tr}\left(\Delta \cdot \Delta^{\dagger}\right)}$. 

Our goal is to produce two matrices $F$ and $G$, with $m'$ columns and rows, respectively, such that their product equates $\tilde{M}$, that is
\begin{equation} \label{eq:FG}
    \tilde{M} = F \cdot G.
\end{equation}
In a tensor network algorithm, the pair $F,G$ corresponds to adjacent tensors where the bond index connecting them has been truncated (e.g. two adjacent MPS tensors during a two-site update in DMRG). An obvious way to obtain $F$ and $G$ is from the SVD of $M$, by choosing e.g. $F = \tilde{W}$ and $G = \tilde{S} \cdot \tilde{V}$. However, here we are interested in obtaining $F$ and $G$ without resorting to an SVD of matrix $M$. 

Remarkably, the above task can be achieved with the polar decomposition which, as described in the previous appendix, can be implemented using a small set of simple matrix operations: matrix-matrix multiplications and additions, as well as matrix transposition and complex conjugation. Next we describe how.

As a first step, we use the polar decomposition to obtain the isometric and positive semi-definite factors $U$ and $H$ of matrix $M$ in Eq. \eqref{eq:polar}. By construction, $H$ has the singular values $s_{\alpha}$ of $M$ as its eigenvalues. As a second step, we compute the polar decomposition of $H - \mathbbm{1}\delta$. Let $U'$ and $H'$ be resulting the unitary and positive semi-definite factors,
\begin{equation} \label{eq: silly}
    H - \mathbbm{1}\delta = U' \cdot H'.
\end{equation}
In general, the polar decomposition $Z = X\cdot |Z|$ of a Hermitian matrix $Z$ with (real) eigenvalues $z_\alpha$ is given in terms of a unitary matrix $X$ and a positive semi-definite matrix $|Z|$ with very simple structure: both $X$ and $|Z|$ have the same eigenvectors as $Z$; moreover, for the $\alpha$-th common eigenvector, $|Z|$ has as eigenvalue the absolute value $|z_{\alpha}|$ of the corresponding eigenvalues $z_{\alpha}$ of $Z$, whereas $X$ has as eigenvalue $\sigma_{\alpha} = \pm 1$, where the sign is such that $z_{\alpha} = \sigma_{\alpha} |z_{\alpha}|$. In other words, the unitary factor $U'$ in \eqref{eq: silly} must have $m'$ eigenvalues $+1$ (for the $m'$ eigenvectors of $H - \mathbbm{1}\delta$ with positive eigenvalues $s_{\alpha}-\delta > 0$) and the rest of eigenvalues must be $-1$ (for the eigenvectors of $H - \mathbbm{1}\delta$ with negative eigenvalues $s_{\alpha}-\delta < 0$). In particular, we can use $U'$ to define two projectors $P_{+}$ and $P_{-}$ onto the positive and negative subspaces of $H - \mathbbm{1}\delta$ (equivalently, the subspaces of $H$ with $s_{\alpha} > \delta$ and with $s_{\alpha} < \delta$) by
\begin{equation} \label{eq: projectorsP}
    P_{\pm} = \frac{\mathbbm{1} \pm U'}{2},
\end{equation}
and use them in turn to define projections $H_{>\delta}$ and $H_{< \delta}$ of matrix $H$ onto its $s_{\alpha}> \delta$ subspace and $s_{\alpha} < \delta$ subspace, 
\begin{eqnarray}
    H_{>\delta} \equiv P_{+} \cdot H \cdot P_{+},~~~H_{>\delta} \equiv P_{-} \cdot H \cdot P_{-},
\end{eqnarray}
such that $H = H_{>\delta} + H_{< \delta}$. We can thus write
\begin{eqnarray}
    M = U \cdot H = U \cdot H_{>\delta} + U \cdot H_{< \delta},
\end{eqnarray}
where the first term $U \cdot H_{>\delta} = U \cdot P_{+} \cdot H$ corresponds to the largest $m'$ singular values $s_{\alpha}$ of $M$. In other words, we have obtained the best rank-$m'$ approximation $\tilde{M}$ to $M$
\begin{equation}
    \tilde{M} = U \cdot P_{+} \cdot H.
\end{equation}
However, we have not yet reduced the number of columns of $U$. For that purpose, we must find an isometry $C_{+}$ with $m'$ columns such that
\begin{equation}
    P_{+} = C_{+}C_{+}^{\dagger}.
\end{equation}
That is, we need to find an orthonormal basis for the $m'$-dimensional column space of the rank-$m'$ projector $P_{+}$. We achieve this using a slight modification of the standard, QR-based subspace-iteration method to avoid the use of a QR decomposition (see Section \ref{sec:subspace}). Then we have
\begin{equation}
\tilde{M} = U \cdot P_{+} \cdot H =  U\cdot C_{+} \cdot C^{\dagger}_{+} \cdot H = \tilde U \cdot \tilde H,
\end{equation}
where matrices
\begin{equation}
    \tilde U \equiv U\cdot C_{+},~~~~\tilde H \equiv C^{\dagger}_{+} \cdot H,
\end{equation}
have $m'$ columns and rows, respectively, and therefore qualify as matrices $F$ and $G$ in the truncated decomposition \eqref{eq:FG}.

A similar approach based on the McWeeny iteration can be used to truncate to a fixed number of singular values, instead of truncating singular values below a
certain threshold $\delta$ \cite{canonical_purify}.

\begin{figure}
\begin{algorithm}[H]
    \caption{Subspace iteration for an $m\times m$ projector matrix $P$ with rank-$m'$ }\label{alg:subspace}
    \begin{algorithmic}[1]
    \Function{subspace($P$)}{}
    \State $m' = \Tr{P}$
    \State $C = \rm{RANDOM}$ $(m,m')$ \Comment{Initial guess of shape $m \times m'$}
    \State $X = P@C$ \Comment{use e.g. SUMMA in distributed setting}
    \State $C=\rm{POLAR\_FACTOR}$ $(X)$\Comment{polar decomposition instead of QR decomposition}
    \State \Return $C$
    \EndFunction
    \end{algorithmic}
\end{algorithm}
\end{figure}

\section{subspace iteration}\label{sec:subspace}

Consider a $m\times m$ Hermitian matrix $P$ that is a rank-$m'$ projector, namely such that 
\begin{equation}
    P \cdot P = P,~~~~\mbox{tr} (P) = m',
\end{equation}
where we also assume that $P$ is rank deficient, meaning $m' < m$.
Our goal is to find an isometric matrix $C$ of shape $m \times m'$ such that we can write $P$ as the product 
\begin{equation}
    P = C \cdot C^{\dagger}.
\end{equation}
For that purpose we can use Alg.~\ref{alg:subspace}. It is a specialization (for a rank-deficient projector $P$) of the subspace iteration method that can more generally be used to compute the first $n$ dominant eigenvectors of a matrix. Specifically, we modified the standard subspace iteration method in two ways:  (1) since $P^2 = P$, a single iteration is sufficient (so we skip looping over steps 4 and 5); (2) we use a polar decomposition (easier to implement on a distributed TPU setting) instead of the usual QR decomposition.

\end{document}